\documentstyle[aps,preprint,tighten,epsf]{revtex}

\def\beq{\begin{equation}}
\def\eeq{\end{equation}}
\def\bea{\begin{eqnarray}}
\def\eea{\end{eqnarray}} 
\def\beqa{\begin{equation}\begin{array}{l}}
\def\eeqa{\end{array}\end{equation}}
\def\eqlab#1{\label{eq:#1}}
\def\figlab#1{\label{fig:#1}}

\def\eref#1{(\ref{eq:#1})}
\def\Eqref#1{Eq.~(\ref{eq:#1})}

\def\Figref#1{Fig.~\ref{fig:#1}}

\def\slap{p \hspace{-2mm} \slash}

\def\half{\mbox{\small{$\frac{1}{2}$}}}

\def\third{\mbox{\small{$\frac{1}{3}$}}}

\def\al{\alpha}
\def\be{\beta}
\def\ga{\gamma} \def\Ga{{\it\Gamma}}
\def\de{\delta} \def\De{\Delta}
\def\veps{\varepsilon}  \def\eps{\epsilon}

\def\la{\lambda} \def\La{{\it\Lambda}}

\def\si{\sigma}

\def\pa{\partial}
\def\vrho{\varrho}

\def\pa{\partial}

\def\nn{\nonumber}
\def\ol#1{\overline{#1}} 

\def\CF#1#2#3#4{#1 {\bf #2}, #4 (#3)}  

\def\prev {Phys. Rev.}

\def\lag{{\mathcal L}}

\begin{document}

\draft

\title{Correspondence of consistent and
inconsistent spin-3/2 couplings via the
equivalence theorem}

\author{V. Pascalutsa}
\address{Department of Physics, 
Flinders University,
Bedford Park, 
SA 5042, Australia}

\date{December 12, 2000}
\maketitle
\thispagestyle{empty}

\begin{abstract}
The Rarita-Schwinger theory of free massive spin-3/2 field
obeys the physical degrees of freedom counting, and one
can distinguish ``consistent'' couplings which
maintain this property and ``inconsistent'' couplings
which destroy it.  We show how one can obtain consistent
couplings from inconsistent ones by a redefinition of the
spin-3/2 field. The field redefinition gives rise to
additional ``contact terms'' which, via the equivalence
theorem, can be associated
with the contribution of the extra lower-spin degrees
of freedom involved by the inconsistent coupling.
\end{abstract}

\bigskip
\bigskip
\bigskip

Relativistic description of hadrons and their dynamics 
often faces with the problem of higher-spin (HS) fields.
A covariant HS field has more components than is
necessary to describe the spin degrees
of freedom (DOF) of the physical particle, and therefore in 
formulating the action certain symmetries must be imposed
to reduce the number of DOF to the physical value.
So any free massless HS action must be
gauge symmetric~\cite{freeHS}, leaving only 2 spin DOF
as is appropriate for a massless particle with spin.
The mass term should (partially) break the gauge symmetries 
such that the number of DOF becomes  $2s+1$. 

Consequently, not every form of the interaction will be
consistent with the free theory construction. The ``inconsistent''
interaction will violate the DOF counting by involving the 
redundant lower-spin sector of the HS field. This gives rise
to the ``lower-spin background'' contributions in observables,
and, in some cases, to the presence of negative norm states 
\cite{Johnson:1961vt} and acausal modes \cite{Velo:1969}.
Many general forms of the interaction are ruled out due
to exhibiting some of these pathologies 
(see {\em e.g.} \cite{Aragone:1980bm,Hagen:1982ez,Deser:2000dz}).

In recent work, focusing on the spin-3/2 field of the $\De (1232)$
isobar, we have argued~\cite{Pas98,PaT99} that the 
interactions with the same 
type of gauge-symmetry as  the kinetic term of the spin-3/2 
field do not change the DOF content of  the free theory, 
hence are ``consistent''; constructed {\em gauge-symmetric} 
$\pi N \De$ and $\ga N\De$ couplings; and shown that they give 
rise to purely spin-3/2 contributions.
On the other hand, during the past three decades, 
a vast number of studies has been done, see {\em e.g.}~\cite{ELA},
using the conventional $\De$ couplings which are inconsistent from the
viewpoint of the DOF-counting. Contributions
due to the spin-1/2 sector of the $\De$ field, so-called
``spin-1/2 backgrounds'', play in these
studies an important, sometimes crucial role while
their physical significance is unclear.

In this Letter we find the correspondence between the
``inconsistent'' and the ``consistent'' (gauge-invariant) couplings.
The two can be related by a redefinition of the spin-3/2 field.
 The redefinition gives also rise to
some higher-order, in the coupling constant, interactions 
(``contact terms'').
As the latter are taken into account, the two theories --- the
inconsistent and the corresponding consistent one --- 
lead to the same $S$-matrix elements (observables)
in accordance with the equivalence theorem~\cite{ET,Coleman:1969sm}.

Thus, on one hand, we can establish the connection between
the studies based on gauge and conventional $\De$ couplings.
On the other hand, the spin-1/2 sector involved by the inconsistent 
coupling can be represented via certain contact interactions,
and depending on the framework one might find a better
physical interpretation of the ``spin-1/2 backgrounds''.
For instance, it is easy to understand how at the {\em tree level} 
the spin-1/2 sector of the $\De$ exchange can
be reabsorbed into some meson and spin-1/2 baryon exchange
contributions~\cite{Pascalutsa:2000bs}, or into some of the 
``low-energy constants'' of ChPT~\cite{Tang:1996sq,Hemmert:1998ye}.
Present equivalence theorem extends to any linear coupling of the spin-3/2
field and to any number of loops.

We
begin with the Lagrangian of the free massive spin-3/2
Rarita-Schwinger (RS) field\footnote{Our conventions: metric tensor
$\eta^{\mu\nu}={\rm diag}(1,-1,-1,-1)$; $\ga$-matrices
$\ga^\mu$, $\ga^5=i\ga^0\ga^1\ga^2\ga^3$,
$\{\ga^\mu,\ga^\nu\}
=2\eta^{\mu\nu}$;
fully antisymmetrized products of $\ga$-matrices 
$\ga^{\mu\nu}=\half[\ga^\mu,\ga^\nu]=\ga^\mu\ga^\nu-\eta^{\mu\nu}$,
$\ga^{\mu\nu\al}=\half \{\ga^{\mu\nu},\ga^\al\}=
i\veps^{\mu\nu\al\be}\ga_\be\ga_5$,
$\ga^{\mu\nu\al\be}=\half [\ga^{\mu\nu\al},\ga^\be]=
i\veps^{\mu\nu\al\be}\ga_5$; spinor indices are omitted. Other
forms of the spin-3/2 Lagrangian, obtained from \eref{rs} 
by redefinition: 
$\psi^\mu  \rightarrow 
(\eta^{\mu\nu} - \half [1+A]\ga^\mu\ga^\nu) \psi_\nu $, involving a parameter
$A$,
are often  used in the literature: 
{\em e.g.}~\cite{Johnson:1961vt,NEK71,Hagen:1971}; for $A \neq -\half$
this redefinition leads to equivalent Lagrangians, and any of them
could be adopted without affecting our results.} 
\beq
\lag_{\rm RS} = \ol\psi_\mu(x)\, \La^{\mu\nu}(i\pa)\, \psi_\nu(x)\, ,
\,\,\,\,
\La^{\mu\nu}(i\pa)\equiv 
\ga^{\mu\nu\al}\,i\pa_\al - 
m\,\ga^{\mu\nu}\, .
\eqlab{rs}
\eeq
Corresponding field equations are
\beq
\eqlab{EOM}
\La^{\mu\nu}(i\pa)\, \psi_\nu = 0 = 
\ga_\mu \La^{\mu\nu}(i\pa)\, \psi_\nu 
=\pa_\mu\La^{\mu\nu}(i\pa)\, \psi_\nu\,.
\eeq
or, equivalently, $(i\ga\cdot\pa -m ) \psi_\mu=0=
\ga\cdot\psi=\pa\cdot\psi$. The kinetic term is, up to
a total derivative, invariant under the gauge transformation: 
\beq
\eqlab{gt}
\psi_\mu(x)\rightarrow \psi_\mu(x)+\pa_\mu\eps(x), 
\eeq
where $\eps(x)$ is a spinor. 

The covariant propagator of the RS field in the momentum
space takes the well-known form:
\beq
\eqlab{rsprop}
S^{\mu\nu}(p)=\frac{\slap+m}{p^2-m^2+i\varepsilon}
\left[-\eta^{\mu\nu}+\third\ga^\mu\ga^\nu
+\frac{1}{3m}(\ga^\mu p^\nu -\ga^\nu p^\mu)
+ \frac{2}{3m^2} p^\mu p^\nu\right],
\eeq
and satisfies
\beq
\eqlab{inverse}
S^{\mu\al}(p)\,\La^{\be\nu}(p)\,\eta_{\al\be}
=\La^{\mu\al}(p)\, S^{\be\nu}(p)\,\eta_{\al\be}
= \eta^{\mu\nu}\,.
\eeq

In considering the interactions, we  will
focus on the {\em linear coupling} of the spin-3/2
field, {\em i.e.}, 
\beq
\lag_{\rm int} = g\,\ol\psi_\mu\,j^\mu + g\,\ol j^\mu \,\psi_\mu \,,
\eeq 
where $j$ can depend on fields other than $\psi$; 
$g$ is a coupling constant.
Consistency with DOF-counting would impose a condition
on $j^\mu$. For example, if the coupling is to be
symmetric under the gauge transformation \eref{gt}, then
$j$ must be divergenceless: $\pa\cdot j =0$.
Suppose, however, that our $j$ does not obey any such condition
and the coupling is {\em inconsistent}.

We can make a field redefinition:
\beq
\eqlab{trans}
\psi_\mu(x)\rightarrow\psi_\mu(x)+g\,\xi_\mu(x)\,, 
\eeq
which gives rise to a new linear coupling $\lag_{\rm int}'$
plus a quadratic (in the coupling constant) interaction $\lag_{C}$:
\bea
 \lag_{\rm RS}+ \lag_{\rm int} & \rightarrow &
\lag_{\rm RS}+ \lag_{\rm int}' + \lag_{C} \,, \nn \\
\lag_{\rm int}' &=& g\,\ol\psi\cdot (j + \La\cdot\xi) 
+ \mbox{H.c.} \equiv g\,\ol\psi\cdot j' + \mbox{H.c.,}\\
\lag_{C}&=&g^2\,[ \ol\xi\cdot \La\cdot\xi + \ol \xi\cdot j
+\ol j\cdot \xi] \,,\nn
\eea
and find that field $\xi_\mu$
can always be chosen such that the new linear coupling 
is {\em consistent}. Namely, we take
\beq
\xi_\mu= (m\,\ga^{\mu\nu})^{-1} j^\nu=
 -\frac{1}{m} {\cal O}^{(-1/3)}_{\mu\nu} j^\nu
\eeq
where ${\cal O}^{(x)}_{\mu\nu}\equiv \eta_{\mu\nu}+x\ga_\mu\ga_\nu$. 
Then
\beq
{j'}^{\mu}=\ga^{\mu\nu\al}i\pa_\al\xi_\nu
\eeq
and consistency condition $\pa\cdot j'=0$ is explicitly obeyed.

Note that in this case $\xi$ and hence $\lag_{C}$ are independent
 of $\psi_\mu$. Therefore, we state that 
(i) an inconsistent linear coupling
of a massive spin-3/2 can in general be transformed, by a redefinition
of the spin-3/2 field, into a consistent coupling
plus an additional quadratic coupling that does not involve
the spin-3/2 field.
Furthermore, if both the Lagrangian and the field transformation
satisfy the conditions of the equivalence theorem~\cite{Coleman:1969sm},
then (ii) the description in terms of $\lag_{\rm int}$ or
$\lag_{\rm int}' + \lag_{C}$ are equivalent at the level
of $S$-matrix.

Moving the second-order coupling $\lag_{C}$ to the other side of the
equation, we obtain a corollary of statements (i) and (ii):  
given any inconsistent linear coupling we can find the supplementary 
second-order interaction which will provide us with the 
description of observables identical to the one with the consistent
coupling. 
The contact interaction $\lag_{C}$ and the spin-1/2 sector raised
by the inconsistent coupling can be put in a direct correspondence. 

\medskip
To demonstrate these statements we focus on the
spin-3/2 coupling to a spin-0 and a spin-1/2 field.
Such couplings are frequently used in describing the coupling
of the decuplet baryons to the pion and nucleon. 
In particular,
the conventional $\pi N\De$ coupling reads~\cite{NEK71}:
\beq
\eqlab{conv}
\lag_{\pi N \De} = g \ol\psi_\mu^{\,i}\,(\eta^{\mu\nu} + z\ga^\mu\ga^\nu)
\, T^a_{ik} \Psi^k \pa_\nu\phi^a + \mbox{H.c.},
\eeq
where $g$ is a dimensionfull coupling constant, and $z$ is an
 ``off-shell parameter''.
Herein we have retained the isospin since it plays some 
role in what follows. 
The pseudo scalar fields $\phi^a$,  correspond
to the pion $\pi^a=(\pi^+,\pi^-,\pi^0)$; spinors $\Psi^k $
correspond to the nucleon $N^k=(p,n)$; the RS fields
$\psi_\mu^i $ represent the $\De^i=(\De^{++},\De^+,\De^0,\De^-)$;
$T^a$ denotes the isospin 1/2 to 3/2
transition matrices satisfying $T^{\dagger a}T^b=\frac{2}{3}\de^{ab}
-\third i\veps^{abc}\tau^c$; $\tau^c$ are the isospin Pauli matrices. 

This is an ``inconsistent'' spin-3/2 coupling.
It violates the DOF counting of the massive RS theory 
for any value of the off-shell parameter $z$. 
Only for the case  of neutral particles (no isospin) 
and $z= -1$ the DOF-counting is correct~\cite{NEK71}.
\footnote{In that case one finds merely 
the Johnson-Sudarshan and Velo-Zwanziger problems \cite{Hagen:1971}.
However, these latter problems will not be addressed here
since we use the ``naive'' Feynman rules (as is usually done!),
hence ignoring the field-dependent determinants in the path-integral
measure which arise in the constrained quantization procedure~\cite{Pas98}
and which contain the problem. In this usual ``naive'' interpretation,
coupling~\eref{conv} for neutral particles and $z=-1$ is ``consistent''.}

To find a corresponding consistent coupling we
make the field transformation \eref{trans}
with 
\beq
\xi_\mu= -\frac{1}{m} {\cal O}^{(-1/3)}_{\mu\vrho} 
{\cal O}^{(z)\vrho\nu} T^a \Psi \pa_\nu\phi^a
\eeq
and thus obtain a gauge-invariant linear coupling
(exactly the one found earlier in \cite{Pas98}):
\beq
\eqlab{GI}
\lag_{\pi N\De}' = -\frac{ig}{2m}\ol G_{\mu\nu} \ga^{\mu\nu\la} T^a \Psi\,
\pa_\la\phi^a + \mbox{H.c.}\, ,
\eeq
where $G_{\mu\nu}=\pa_\mu\psi_\nu-\pa_\nu\psi_\mu$, plus the
second-order term:
\beq
\lag_{\pi\pi NN}= -\left(\frac{g}{m}\right)^2
\ol\Psi\,{\cal O}^{(x)}_{\vrho\mu}\, (\ga^{\mu\nu\al}\,i\pa_\al +
m\,\ga^{\mu\nu})\,{\cal O}^{(x)}_{\nu\si}\,T^{\dagger b}T^c \Psi\, 
(\pa^\vrho\phi^{\dagger b}) (\pa^\si\phi^c)\,.
\eeq
with $x=-\third (1+z)$. Thus, $\lag_{\rm RS}+ \lag_{\pi N\De}
\rightarrow 
\lag_{\rm RS}+ \lag_{\pi N\De}' + \lag_{\pi\pi NN}$. 

To check that the field transformation leaves the
$S$-matrix invariant let us first consider some simplest 
matrix elements involving the two vertices:
\bea
&&\Ga^{\mu\, a} (k) \equiv \Ga^\mu (k)\,T^a,\,\,\,
\Ga^\mu (k)= g\,(\eta^{\mu\nu}+z\ga^\mu\ga^\nu) k_\nu
\,\,\,\, \mbox{(inconsistent)}\nn\\
&&\tilde\Ga^{\mu\, a} (k,p) \equiv \tilde\Ga^\mu (k,p)\,T^a,\,\,\,
\tilde{\Ga}^\mu (k,p)=-(g/m)\,\ga^{\mu\nu\al}k_\nu p_\al 
\,\,\,\, \mbox{(consistent)}\, .\nn
\eea

The $\De$ production amplitude is apparently
the same for both vertices
\beq
\ol u(p')\,\Ga^{\mu\, a} (p'-p)\, u_\mu (p) 
=\ol u(p')\,\tilde\Ga^{\mu\, a} (p'-p,p)\, u_\mu (p)
=g\,\ol u(p')\,(p'-p)^\mu\, u_\mu (p)\,T^a,
\eeq
where $u(p)$ is the nucleon spinor, $u_\mu (p)$ is the 
free RS vector-spinor
satisfying $(\slap - m) u_\mu=0=p\cdot u =\ga\cdot u$,
for $p^2=m^2$.

However for the $\De$-exchange amplitudes in pion-nucleon scattering,
\Figref{exchange},
the two vertices yield quite different results. 
The inconsistent coupling
involves the spin-1/2 sector of the RS propagator, and therefore
the exchange amplitude,
\beq
\eqlab{a1}
\Ga^{\mu}(k') S_{\mu\nu}(p+k) \Ga^\nu (k) \sim
\frac{g^2}{m-(p+k)\cdot\ga} P^{3/2}_{\mu\nu}(p+k)\, {k'}^\mu k^\nu +
\mbox{``spin-1/2 background''}\,,
\eeq 
contains the controversial spin-1/2 background contributions,
in addition to the spin-3/2 propagation represented
by the spin-3/2 projection operator: 
\beq
P^{3/2}_{\mu\nu}(p)=\eta_{\mu\nu} - \frac{1}{3}\gamma_\mu\gamma_\nu
     - \frac{1}{3p^2} (p\hspace{-1.65mm}\slash\gamma_\mu p_\nu
        + p_\mu\gamma_\nu p\hspace{-1.65mm}\slash). 
\eeq
In contrast, the consistent 
coupling,
because of the property $p\cdot \tilde\Ga(k,p)=0$, gives rise to
only the spin-3/2 propagation \cite{Pas98,PaT99}, namely
\beq
\eqlab{a2}
\tilde{\Ga}^{\mu}(k',p_s) S_{\mu\nu}(p_s) \tilde{\Ga}^\nu (k,p_s) =
\frac{g^2}{m-\slap_s} \frac{p_s^2}{m^2} P^{3/2}_{\mu\nu}(p_s)\, 
{k'}^\mu k^\nu\,.
\eeq 
where $p_s=p+k=p'+k'$.

Because the decomposition into the pure spin-3/2 and
spin-1/2 sector is nonlocal (the projection operators
are singular at $p^2=0$), it is not at all obvious that
the difference between the amplitudes \eref{a1} and \eref{a2}
can be compensated by the local contact term $\lag_{\pi\pi NN}$.
However it indeed happens, as can most easily be seen
for the case $z=-1$, when
\beq
\tilde{\Ga}^\mu(k,p)-\Ga^\mu(k)=-(g/m) \La^{\mu\nu}(p)\,k_\nu\,.
\eeq
Using this identity and \Eqref{inverse} we find for
the $s$-channel exchange the difference between the 
amplitudes \eref{a1} and \eref{a2} is
\bea
M_C^{(s)}&=&[\tilde{\Ga}(k',p_s)\cdot S(p_s)\cdot \tilde{\Ga} (k,p_s)
-\Ga (k') \cdot S_{\mu\nu}(p_s)\cdot \Ga (k)]\, T^{\dagger a} T^b \nn\\
&=&  \left[ -(g/m)\, (k'\cdot \Ga(k)+ \Ga(k')\cdot k)\,+ 
(g/m)^{2} k'\cdot\La(p_s)\cdot k\right] \, T^{\dagger a} T^b.
\eea
A similar contribution comes from the
u-channel graph such that the total difference sums up into
\beq
M_C^{(s)}+M_C^{(u)}=-(g/m)^{2}\,( T^{\dagger a} T^b-T^{\dagger b} T^a) 
[ \half\ga^{\mu\nu\al}\,(p+p')_\al +
m\,\ga^{\mu\nu}] k_\mu' k_\nu\,,
\eeq
which is exactly canceled by the contact 
interaction $\lag_{\pi\pi NN}$.\footnote{Notably, for the case of
{\em neutral pions}, obtained by neglecting the
isospin complications and taking the real scalar field in
\Eqref{conv}, the contact term $\lag_{\pi\pi NN}$ vanishes.
As we have already remarked, for this case, coupling~\eref{conv}
with $z=-1$ is consistent with the DOF-counting~\cite{NEK71}, does not
involve the spin-3/2 components, and hence needs no
additional terms to match the corresponding gauge-invariant
coupling.}

Clearly, Green's functions which do not represent
observable quantities need not be the same. For
instance, the one-loop $\De$ self-energy
will be different for the two couplings. However,
at the level of the $S$-matrix the equivalence
is restored. That is, the amplitude containing
the self-energy with the consistent coupling 
(l.h.s.\ in \Figref{loop}) is identical to the
one-loop amplitude with the inconsistent coupling
plus the contact term (r.h.s.\ in \Figref{loop}).
It is easy to convince oneself that
this equivalence will persist to any 
number of loops.

In the analysis of the $\pi N$ scattering
we have recently noted numerically~\cite{Pascalutsa:2000bs}
that, once the $\rho$ and $\si$ meson exchange are included,
the gauge-invariant \Eqref{GI} and conventional \Eqref{conv}  
couplings, at the tree level, give the same prediction for the $\pi N$ 
threshold parameters provided some coupling constants are readjusted.
This observation can now be understood as the very low-energy meson
exchanges may take the role of the contact term $\lag_{\pi\pi NN}$
needed to provide the equivalence between the couplings. 

Another relevant statement with respect to the $\pi N\De$ coupling
has been recently made by 
Tang and Ellis~\cite{Tang:1996sq} from the point of view of 
Chiral Perturbation Theory (ChPT). They have shown
that the contribution of the ``off-shell parameter''
to the $\De$-exchange amplitude can be absorbed into
contact terms which are already present in the ChPT Lagrangian.
Therefore, they argue, this parameter is redundant in ChPT.
By using the field redefinition we can confirm their result
and extend it to any linear spin-3/2 couplings, {\em e.g.} the
$\ga N \De$ couplings.

It is further possible to argue that 
within the ChPT framework any
linear spin-3/2 coupling is acceptable:
in ChPT the additional $\lag_C$ type of terms,
which provide the equivalence of ``inconsistent'' and 
``consistent'' couplings, are to be included anyway with arbitrary
coefficients and in both situations. 
(Thus, the effective
Lagrangian with an inconsistent coupling and a $\lag_C$ term
with arbitrary coefficient $c_1$, is equivalent to
the Lagrangian with a consistent coupling and $\lag_C$ term
with a different but yet arbitrary coefficient $c_2=c_2(c_1,g/m)$.
The spin-1/2 background will lead only to a
change in the values of some coupling constants.) 

Nevertheless, let us emphasize the use of the gauge-invariant
couplings makes the calculations much easier and more transparent.
In particular, the spin-1/2 sector can be entirely dropped from
the RS propagator, while analyzing
the spin-3/2 self-energy,  one does not
need to consider the ten scalar functions of
the most general tensor structure~\cite{complete_se}, but only two of 
them~\cite{Kondratyuk:2000xg,Pascalutsa:2000bs}, 
just as in the spin-1/2 case. 

Besides the technical advantages, gauge-invariant couplings involve 
the physical highest-spin contributions only, and hence they
are preferable in the analysis of separate contributions and effects
due to spin-3/2 particles versus the rest. This is important when
the properties of separate HS resonances
are being extracted in a model-dependent way from experimental data. 
Here we in particular keep in mind the current programs at CEBAF and MAMI
aimed to measure various electromagnetic properties of the 
$\De$(1232) isobar.

Much of the said about the linear couplings
is applicable to the {\em quadratic}, etc., couplings as well. Given
any inconsistent coupling of a massive RS field
$\psi_\mu$, we can obtain an on-shell equivalent
consistent coupling by the replacement:
\beq
\psi_\mu \rightarrow \frac{i}{m} {\cal O}^{(-1/3)}_{\mu\la}\,\ga^{\la\al\be}
\,\pa_\al\psi_\be \,.
\eeq
It is then possible to work out the exact field
transformation relating the couplings and the 
supplementary higher-order terms providing
their equivalence at the $S$-matrix level.
Working these out for any specific example
is beyond the scope of this letter. Without
going into details it is already clear that
once the coupling involves several RS fields,
the transformation must be nonlinear in
the RS field and hence the number of supplementary
terms is infinite.    


In conclusion, we have shown that for any linear coupling
of a massive spin-3/2 field all the possibly arising spin-1/2
contributions can be represented by a single interaction term
independent of the spin-3/2 field itself. This is proven
by showing that any ``inconsistent'' coupling involving 
the spin-1/2 components
can be transformed by a field redefinition into a ``consistent''
gauge-invariant
coupling that decouples the spin-1/2 sector. The additional
interaction term arises in the course of field transformation
and is needed to provide the equivalence of the two couplings
at the level of $S$-matrix elements.
We emphasize that consistent interactions have 
conceptual and technical advantages, making the calculations simpler
and with more transparent interpretation.

\acknowledgments
I am very thankful to Jambul Gegelia for many interesting discussions,
and to Iraj Afnan, Stanley Deser, Thomas Hemmert, Justus Koch, Andrew
Waldron for valuable remarks.
The work is supported by the Australian Research Council (ARC).


\begin{figure}[t]
\begin{center}
\epsffile{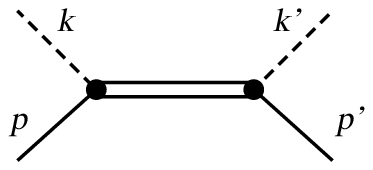}
\end{center}
\caption{The $s$-channel $\Delta$-exchange in $\pi N$ scattering. 
The dashed, solid, and double lines denote the pion, the nucleon, 
and the $\Delta$, respectively.}
\figlab{exchange}
\end{figure}

\begin{figure}[t]
\begin{center}
\epsffile{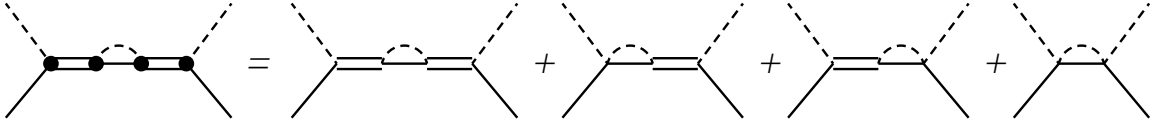}
\end{center}
\caption{The equivalence of a $\pi N$-scattering loop amplitude with 
the consistent $\pi N\De$ coupling (denoted by a dot)
on the l.h.s., and in the corresponding
inconsistent model on the r.h.s.; crossed graphs are omitted.}
\figlab{loop}
\end{figure}

\end{document}